\documentclass[reprint, superscriptaddress, 
[floatfix, longbibliography, 
amsfonts, amssymb, amsmath]{revtex4-1}

\usepackage{amsmath}
\usepackage[]{graphicx}
\usepackage{bm}
\usepackage{dcolumn}
\usepackage{color}
\usepackage{multirow}
\usepackage{ulem} 

\usepackage[
colorlinks=true, citecolor=blue, urlcolor=blue, linkcolor=blue,
setpagesize=false
]{hyperref}

\begin{document}
\title{Dirac electrons in the square-lattice Hubbard model 
with a $d$-wave pairing field: 
The chiral Heisenberg universality class revisited
} 

\author{Yuichi Otsuka}
\email{otsukay@riken.jp}
\affiliation{Computational Materials Science Research Team, 
RIKEN Center for Computational Science (R-CCS), 
Kobe, Hyogo 650-0047, Japan}

\author{Kazuhiro Seki}
\affiliation{Computational Quantum Matter Research Team, 
RIKEN Center for Emergent Matter Science (CEMS), 
Wako, Saitama 351-0198, Japan}

\author{Sandro Sorella}
\affiliation{Computational Materials Science Research Team, 
RIKEN Center for Computational Science (R-CCS), 
Kobe, Hyogo 650-0047, Japan}
\affiliation{SISSA -- International School for Advanced Studies, 
Via Bonomea 265, 34136 Trieste, Italy}
\affiliation{Democritos Simulation Center CNR--IOM Instituto Officina 
dei Materiali, Via Bonomea 265, 34136 Trieste, Italy}

\author{Seiji Yunoki}
\affiliation{Computational Materials Science Research Team,
RIKEN Center for Computational Science (R-CCS), 
Kobe, Hyogo 650-0047, Japan}
\affiliation{Computational Quantum Matter Research Team, 
RIKEN Center for Emergent Matter Science (CEMS), 
Wako, Saitama 351-0198, Japan}
\affiliation{Computational Condensed Matter Physics Laboratory,
RIKEN,
Wako, Saitama 351-0198, Japan}

\date{\today}

\begin{abstract}
 We numerically investigate the quantum criticality of 
 the chiral Heisenberg universality class 
 with the total number of fermion components $N=8$
 in terms of the Gross-Neveu theory.
 Auxiliary-field quantum Monte Carlo simulations are performed 
 for the square lattice Hubbard model 
 in the presence of a $d$-wave pairing field,
 inducing Dirac cones in the single-particle spectrum.
 This property makes the model particularly interesting because   
 it turns out to belong to the same universality class of 
 the Hubbard model on the honeycomb lattice,
 which is the canonical model for graphene,
 despite the unit cells being apparently different 
 (e.g.,  they contain one and two sites, respectively).
 We indeed show 
 that the two phase transitions,
 expected to occur on the square and on the honeycomb lattices,
 have the same quantum criticality.
 We also argue that 
 details of the models, i.e.,
 the way of counting $N$ and the anisotropy of the Dirac cones,
 do not change the critical exponents.
 The present estimates of the exponents 
 for the $N=8$ chiral Heisenberg universality class are 
 $\nu=1.05(5)$, $\eta_{\phi}=0.75(4)$, and $\eta_{\psi}=0.23(4)$,
 which are compared with the previous numerical estimations.
\end{abstract}

\maketitle

\section{\label{sec:intro}Introduction}

The study of quantum critical points in Dirac fermions has received
increased attention across many research fields.
Massless Dirac fermions emerge as quasiparticles in various 
condensed-matter systems~\cite{Wehling_2014,Vafek_AnnuRev2014}.
The Coulomb interaction among electrons is almost inevitable 
even within the quasiparticle picture, and so far the most popular  
models adopted to describe 
the electron correlation in the Dirac fermions 
are standard Hubbard models 
of which the kinetic-energy parts are characterized by linear dispersions
near the Fermi level.
Numerical studies on these models have typically found evidence of a Mott transition 
between a semimetal (SM) and an antiferromagnetic (AF)
insulator~\cite{Sorella_EPL1992,Otsuka_PRB2002a,Paiva_PRB2005,Meng_Nature2010,Sorella_SR2012,Otsuka_Proc2013,Otsuka_Proc2014,Ixert_PRB2014,Chen_PRB2014}.
Further analytical studies motivated by graphene~\cite{Herbut_PRL2006,Herbut_PRB2009a,Herbut_PRB2009b,Ryu_PRB2009}
have revealed that an effective theory for the Mott transition is described by 
the Gross-Neveu (GN) model~\cite{Gross_PRD1974},
a well-studied effective model in high-energy physics~\cite{Gracey_NuclPhysB1990,Vasilev_1993,Rosenstein_PhysLettB1993,Karkkainen_NuclPhysB1994,Gracey_1994,Rosa_PRL2001,Hofling_PRB2002}.
Since it had been predicted by the GN theory that
the phase transitions of the interacting Dirac fermions are classified into 
three universality classes, 
namely chiral Ising, chiral $XY$, and chiral Heisenberg classes,
corresponding to the Z$_{2}$, U(1), and SU(2) symmetry breaking,
a number of quantum Monte Carlo (QMC) calculations have been performed
on various lattice models realizing quantum phase transitions in the Dirac fermions~\cite{Wang_NJP2014,Li_NJP2015,Wang_PRB2015,Hesselmann_PRB2016,Wang_PRB2016,Xu_arXiv2019,Liu_PRB2020,Huffman_PRD2020,Li_NatCom2017,Otsuka_PRB2018,Xu_PRB2018,Li_PRB2020,Assaad_PRX2013,ParisenToldin_PRB2015,Otsuka_PRX2016,Buividovich_PRB2018,Lang_PRL2019,Liu_Natcomm2019,Ostmeyer_arXiv2020,Xu_arXiv2020}. 
The proliferation of numerical studies on these lattice models has also led to 
a renewed interest in the continuum GN model~\cite{Janssen_PRB2014,Classen_PRB2017,Zerf_PRD2017,Knorr_PRB2018,Gracey_PRD2018,Ihrig_PRB2018}.

Since a universality class is characterized and distinguished by a set of critical exponents, 
the evaluation of the critical exponents  is of general interest,
and particularly challenging for theorists.
In spite of considerable efforts devoted to determine the critical exponents 
of the GN universality classes, 
only partial agreement between different theoretical methods has been achieved;
in particular large discrepancies between numerical and analytical approaches
have remained.
Indeed, a satisfactory consistency has been found only for the $N=4$ chiral Ising class
($N$ denotes the total number of fermion components)~\cite{Ihrig_PRB2018,Huffman_PRD2020},
and other classes including the $N=8$ chiral Heisenberg class, 
which is of particular interest because of its tight connection with 
graphene~\cite{Herbut_PRL2006,Herbut_PRB2009a,Herbut_PRB2009b,Ryu_PRB2009},
have to be further explored.

It should also be pointed out that
the critical exponents are not only a subject of 
purely academic interest 
but are also practically useful to identify the phase transitions.
Such examples are found in the studies of
the honeycomb bilayer model~\cite{Pujari_PRL2016},
the Kekul\'{e} valence-bond-solid transition~\cite{Li_NatCom2017,Xu_arXiv2019}, and
the quantum spin Hall insulator transition~\cite{Liu_Natcomm2019}.

In this paper, we investigate the quantum phase transition 
in another celebrated manifestation of  Dirac fermions 
in condensed-matter physics:
the spectrum of a $d$-wave superconductor (SC)~\cite{Scalapino_PRB1986,Lee_PRL1993}, 
showing clear nodal quasiparticles in the single-particle spectrum.
Specifically, we consider an effective square lattice 
model of a $d$-wave SC in the presence of the Hubbard $U$ interaction. 
In this model there are four Dirac cones for each spin component,
and therefore the phase transition triggered by $U$ should in principle 
belong to the $N=8$ chiral Heisenberg class, 
the same universality class of 
the Hubbard model on the honeycomb lattice
(hereinafter referred to as the honeycomb lattice model)~\cite{Sorella_EPL1992,Paiva_PRB2005,Meng_Nature2010,Sorella_SR2012,Otsuka_Proc2013,Ixert_PRB2014,Chen_PRB2014,Assaad_PRX2013,ParisenToldin_PRB2015,Otsuka_PRX2016,Buividovich_PRB2018,Ostmeyer_arXiv2020},
which is closely related to graphene~\cite{Herbut_PRL2006,Herbut_PRB2009a,Herbut_PRB2009b,Drut_PRL2009}.
Thus the same critical exponents are expected as those for the honeycomb lattice model.

This is remarkable because 
the Hubbard model with the $d$-wave pairing field studied here, 
is apparently different from the honeycomb lattice model.
First, the former model has four independent Dirac cones without sublattice
in contrast to two Dirac cones with two sublattices for the latter model,
while the total number of the fermion components is the same, i.e., $N=8$ for both models.
Second, in the $d$-wave SC, the Dirac cone is in general anisotropic, 
because the velocity at the Dirac point
depends on the chosen direction in momentum space. 
From the point of view of renormalization group, 
the relativistic invariance may emerge at the critical point. 
However, the effect of the anisotropy on the quantum criticality 
cannot be studied in the honeycomb lattice model which has the isotropic Dirac cones.
According to the notion of the universality class stating that
the criticality does not depend on the details of the models,
the critical exponents for both models should in principle be the same.
Therefore we expect that our work represents
a nontrivial test of  this universality assumption for the exponents.

The rest of the paper is organized  as follows.
In the next section, the model is defined and the QMC method is briefly explained.
In Sec.~\ref{sec:result}, the results of the QMC simulations are analyzed by 
various methods such as a crossing-point analysis based on 
the phenomenological renormalization argument and a data-collapse method
of the finite-size scaling ansatz.
The obtained critical exponents are discussed 
in comparison with the previous estimations, 
before concluding the paper, in Sec.~\ref{sec:conclusion}.

\section{Model and Method}\label{sec:model-method}

\subsection{Model}\label{subsec:model}

We study the two-dimensional Hubbard model at half filling
with the $d$-wave BCS SC order parameter 
described by the following Hamiltonian:
\begin{equation}
 H = H_{\mathrm{BCS}} + H_{U}, \label{eq:model}
\end{equation}
where
\begin{equation}
 H_{\mathrm{BCS}}  
 = 
\sum_{\langle i, j \rangle}
 \left\{
 \begin{pmatrix}
	c_{i \uparrow}^{\dagger} & 	c_{i \downarrow}^{}	
 \end{pmatrix}
 \begin{pmatrix}
  -t_{} & \Delta_{ij} \\
 \Delta_{ij}^{\ast} & t_{}
 \end{pmatrix}
 \begin{pmatrix}
  c_{j \uparrow}^{} \\
  c_{j \downarrow}^{\dagger}	
 \end{pmatrix} 
 + \mathrm{h.c}
 \right\} \label{eq:H_BCS} \\
\end{equation}
and
\begin{equation}
H_{U}  = U \sum_{i} n_{i \uparrow} n_{i \downarrow}. \label{eq:Hubbard}
\end{equation}
Here, $c_{i \sigma}^{\dagger}$ creates an electron with spin $\sigma\,(=\uparrow,\downarrow)$ 
at site $i$ of position $\bm{r}_{i}$
and $n_{i \sigma}=c_{i \sigma}^{\dagger}c_{i \sigma}$ is a number operator.
In the noninteracting part, $H_{\mathrm{BCS}}$,
$t$ represents the transfer integral chosen as an energy unit, i.e., $t=1$,
and $\Delta_{ij}$ is the BCS SC order parameter,
with the sum indicated by $\langle i,j\rangle$ 
running over all pairs of nearest-neighbor sites $i$ and $j$.
We consider the model on a square lattice of linear dimension $L$. 
The BCS order parameter $\Delta_{ij}$ with the $d$-wave symmetry
is set to have a uniform amplitude $\Delta$ between the nearest-neighbor 
sites:
$\Delta_{ij}$ = $\Delta$ ($-\Delta$) for 
sites $i$ and $j$ 
aligned along the $x$ ($y$) direction.
In the interacting part of Eq.~(\ref{eq:Hubbard}),
$U(>0)$ denotes the repulsive interaction, 
which triggers the quantum phase transition 
from the SM~\cite{noteSM} to the AF insulator,
breaking both the chiral and SU(2) symmetries.
Since we take into account the amplitude $\Delta$ as a model parameter,
the U(1) symmetry is explicitly broken from the outset, and thus 
the AF insulator with two Goldstone modes~\cite{Nielsen_NuclPhysB1976,Watanabe_AnnuRev2020} 
coexists with the $d$-wave SC order.
Note that, 
according to a very recent report~\cite{Xu_arXiv2020}, 
the U(1) symmetry plays no important role in the AF transition.

\begin{figure}[tb]
 \centering
 \includegraphics[width=0.49\linewidth]{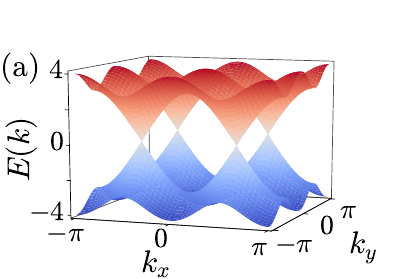}
 \includegraphics[width=0.49\linewidth]{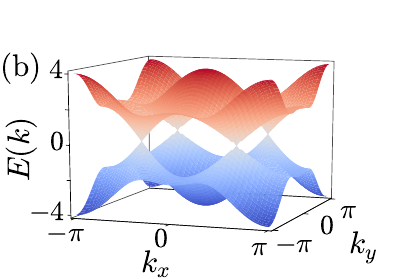}\\
 \includegraphics[width=0.49\linewidth]{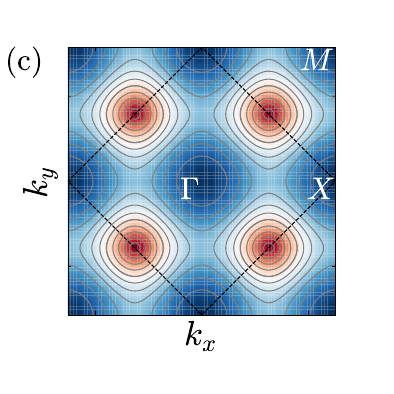}
 \includegraphics[width=0.49\linewidth]{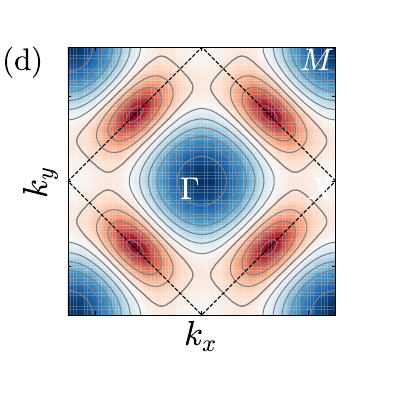}
 \caption{\label{fig:dispersion}%
 The noninteracting energy dispersion $E(\bm{k})$ for 
 (a) $\Delta=1$ and (b) $\Delta=0.5$, and
 the corresponding contour plot of the lower band 
 for (c) $\Delta=1$ and (d) $\Delta=0.5$.
 In panels (c) and (d),
 high symmetric momenta are denoted 
 $\Gamma$ for $\bm{k}=(0,0)$,
 $X$ for $\bm{k}=(\pi,0)$, and 
 $M$ for $\bm{k}=(\pi,\pi)$,
 and 
 the Fermi surface in the case of $\Delta=0$ is indicated by dashed lines.
 }
\end{figure}

The noninteracting Hamiltonian of Eq.~(\ref{eq:H_BCS})
is expressed in  momentum space as follows:
\begin{equation}
 H_{\mathrm{BCS}}
= 
\sum_{\bm{k}}
 \begin{pmatrix}
	c_{\bm{k} \uparrow}^{\dagger} & 	c_{-\bm{k} \downarrow}^{}	
 \end{pmatrix}
 \begin{pmatrix}
  \epsilon_{\bm{k}}        &  \Delta_{\bm{k}}\\
  \Delta_{\bm{k}}^{\ast}   & -\epsilon_{\bm{k}}
 \end{pmatrix}
 \begin{pmatrix}
  c_{ \bm{k} \uparrow}^{} \\
  c_{-\bm{k} \downarrow}^{\dagger}	
 \end{pmatrix},
 \label{eq:hbcs}
\end{equation}
where
\begin{equation}
\epsilon_{\bm{k}} = -2t \left( \cos{k_{x}} + \cos{k_{y}} \right)
\label{eq:band}
\end{equation}
and
\begin{equation}
\Delta_{\bm{k}}  =  2\Delta \left( \cos{k_{x}} - \cos{k_{y}} \right).
\label{eq:delta}
\end{equation}
The energy dispersion of Bogoliubov quasiparticles is then obtained as
$E(\bm{k}) = \pm \left( \epsilon_{\bm{k}}^{2} + |\Delta_{\bm{k}}|^{2} \right)^{1/2}$,
which has four independent Dirac points at $\bm{k}=(\pm \pi/2, \pm \pi/2)$ and $(\pm \pi/2, \mp \pi/2)$
as shown in Fig.~\ref{fig:dispersion}.
Together with the spin degrees of freedom, 
the effective model in the continuum limit is 
the GN model with a total number of fermion components 
$N=8$~\cite{Sachdev_Book1999,Vojta_PRL2000,Khveshchenko_PRL2001},
which is the same as the honeycomb lattice model
or the Hubbard model on the square lattice with $\pi$ flux
(referred to as the $\pi$-flux model in the following)~\cite{Otsuka_PRB2002a,Otsuka_Proc2014,Ixert_PRB2014,ParisenToldin_PRB2015,Otsuka_PRX2016},
although the counting of the fermion components is different:
two Dirac cones, two sublattices, and two spin components
in the latter models.

The Dirac cones described by $H_{\mathrm{BCS}}$ are in general anisotropic,
i.e., elliptic cones, and the ellipticity is determined by $|\Delta/t|$.
To be concrete, let us focus on the low-lying excitations around one Dirac
point at $\bm{k}_{\mathrm{D}}=(\pi/2, \pi/2)$.
The energy dispersion can be expanded 
with a small wave vector 
$\delta \bm{k}=\bm{k}-\bm{k}_{\mathrm{D}}=(\delta k_{x}, \delta k_{y})$ 
as $E(\bm{k}_{\mathrm{D}}+\delta \bm{k}) \approx \pm \mathcal{E}(\delta \bm{k})$,
where
\begin{equation}
 \mathcal{E}(\delta \bm{k}) = 
  \sqrt{
  v_{\mathrm{F}}^{2} \left( \frac{\delta k_{x} + \delta k_{y}}{\sqrt{2}} \right)^{2}
  +
  v_{\mathrm{\Delta}}^{2} \left( \frac{\delta k_{x} - \delta k_{y}}{\sqrt{2}} \right)^{2}
  }
\end{equation}
and the nodal Fermi velocity $v_{\mathrm{F}} \equiv 2 \sqrt{2} t$ 
(the gap velocity $v_{\mathrm{\Delta}} \equiv 2 \sqrt{2} |\Delta|$) 
is the velocity perpendicular (parallel) to the Fermi surface 
of $H_{\rm BCS}$ with $\Delta=0$ at half filling. 
The Dirac cone becomes isotropic if and only if 
$|\Delta/t|=1$, as shown in Fig.~\ref{fig:dispersion}(c), and 
otherwise it is elliptic, as shown in Fig.~\ref{fig:dispersion}(d)~\cite{Otsuka_PRB2002b}.
Therefore, our model is considered as a tunable model,
where the velocity is controllable and 
the Dirac cone can be deformed elliptically
with the Dirac points kept at $\bm{k}=(\pm \pi/2, \pm \pi/2)$ and $(\pm \pi/2, \mp \pi/2)$.
This is a convenient feature, allowing us to study 
the putative universal nature of the
phase transitions based on lattice model simulations.

The quasiparticle density of states $D(E)$ per site
close to the Dirac point is given as 
\begin{equation}
D(E)=\frac{2\pi N_{\mathrm{Dirac}}} {\bar{v}^2 V_{\mathrm{BZ}} } |E|,
\label{eq:DOS}
\end{equation}
where 
$N_{\mathrm{Dirac}}$
is the number of Dirac points in the Brillouin zone (BZ),
$V_{\mathrm{BZ}}$ is the volume of the BZ, and
$\bar{v}$ is the geometric mean of $v_{\mathrm{F}}$ and $v_{\mathrm{\Delta}}$, i.e., $\bar{v}=\sqrt{v_{\mathrm{F}}v_{\mathrm{\Delta}}}$. 
Note that Eq.~(\ref{eq:DOS}) also holds 
for the honeycomb lattice model and the $\pi$-flux model.
Therefore, $\bar{v}$ plays essentially the same role as 
the Dirac Fermi velocity $v_{\mathrm{F}}^0$ in the isotropic model.

\subsection{Method}\label{subsec:method}

The model described by the Hamiltonian $H$ in Eq.~(\ref{eq:model}) is investigated
by the auxiliary-field quantum Monte Carlo (AFQMC) 
method~\cite{Blankenbecler_PRD1981,Hirsch_PRB1985,White_PRB1989,Assaad_2008,Becca_2017}
at half filling,
where the fermionic negative-sign problem does not occur.
An expectation value of a physical observable $O$ at zero temperature is 
calculated for the ground-state wave function
projected from a left (right) trial wave function
$\langle \psi_{\mathrm{L}}|$ ($|\psi_{\mathrm{R}} \rangle$),
\begin{equation}
 \langle O \rangle =
  \frac{
  \langle \psi_{\mathrm{L}}| e^{-\frac{\beta}{2} H} 
  O
  e^{-\frac{\beta}{2} H}  |\psi_{\mathrm{R}} \rangle
  }{
  \langle \psi_{\mathrm{L}}| e^{-\beta H} |\psi_{\mathrm{R}} \rangle
  },
\end{equation}
where projection time denoted by $\beta$ is set to be proportional to $L$
with the Lorentz invariance assumed~\cite{Herbut_PRB2009b,Herbut_PRB2009a}.
A slice of the Suzuki-Trotter decomposition~\cite{Suzuki_1976,Trotter_1959}
is chosen as $\Delta \tau = \beta/M = 0.1$
with $M$ being integer, which is confirmed to result only in negligible 
systematic errors compared with statistical errors in the Monte Carlo sampling.
The discrete Hubbard-Stratonovich transformation~\cite{Hirsch_1984}
is employed to decouple the interaction term $e^{-\Delta \tau H_{U}}$, 
which introduces a real auxiliary Ising field coupled to spin 
at each site in space and imaginary time.
The simulations are performed on 
finite-size lattices of $L=8, 12, 16, 20, 24, 32, 40$
with periodic boundary conditions
for several values of $U$
below and above the quantum critical point $U_{\mathrm{c}}$ of the phase transition.

\section{Results}\label{sec:result}

\subsection{crossing-point analysis}\label{subsec:crossing_point}

\begin{figure}[tb]
 \centering
 \includegraphics[width=1.0\linewidth]{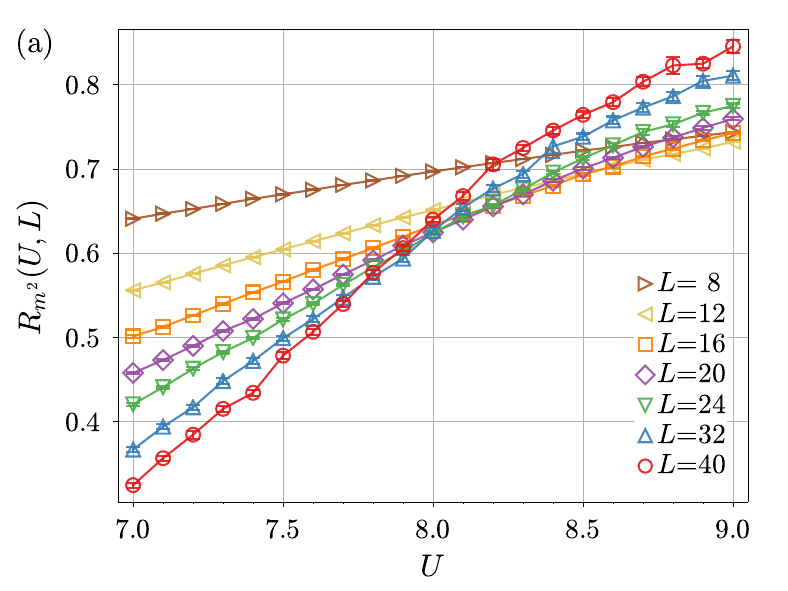}\\
 \includegraphics[width=1.0\linewidth]{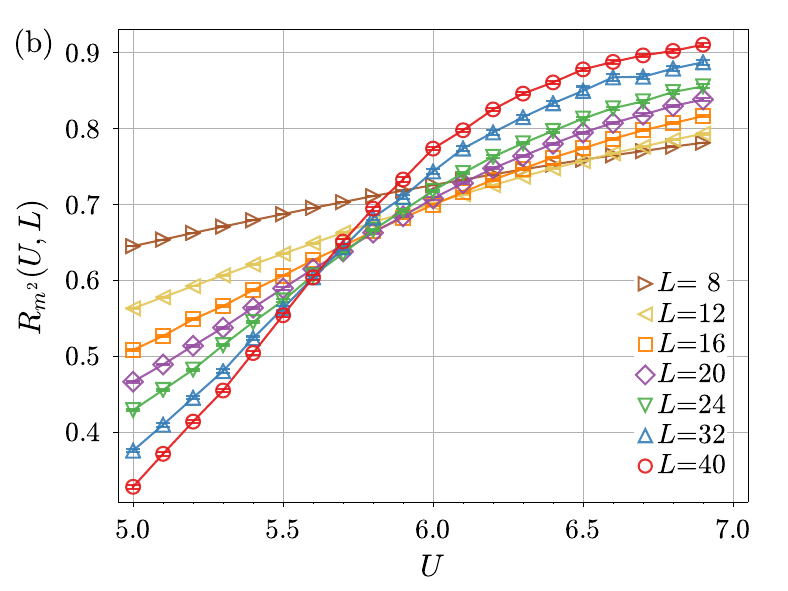}
 \caption{\label{fig:crossing_points}
 Correlation ratio $R_{m^{2}}(U,L)$ of spin structure factor
 as a function of $U$ for various system sizes $L$;
 (a) $\Delta=1$ and (b) $\Delta=0.5$.
 }
\end{figure}

Using AFQMC, we calculate the spin structure factor 
defined as
\begin{equation}
 S(\bm{k}) = 
  \frac{1}{L^{2}} 
  \sum_{i,j}
  e^{ i \bm{k} \left( \bm{r}_{i} - \bm{r}_{j} \right)}
  \langle
  \bm{S}_{i}\cdot \bm{S}_{j}
  \rangle,
  \label{eq:Sk}
\end{equation}
where 
$
\bm{S}_{i} = 
\frac{1}{2}\sum_{s, s^{\prime}}
c_{i s}^{\dagger} 
\left( \bm{\sigma} \right)_{s s^{\prime}} 
c_{i s^{\prime}}
$
is the spin operator
with $\bm{\sigma}=(\sigma_{x}, \sigma_{y}, \sigma_{z})$ being the vector of Pauli matrices.
For large $U$,
$S(\bm{k})$ is peaked at the AF ordering momentum,
$\bm{K}=(\pi, \pi)$.
The critical point $U_{\mathrm{c}}$ of the AF transition can be located by
monitoring the correlation ratio,
\begin{equation}
 R_{m^{2}}(U,L) = 1 - \frac{S(\bm{K}+\bm{b}/L)}{S(\bm{K})},
\end{equation}
where $\bm{b}$ denotes the smallest reciprocal-lattice vector.
In the AF ordered phase, $R_{m^{2}}(U,L)$ scales to 1 in the thermodynamic limit,
as the Bragg peak in $S(\bm{k})$ becomes infinitely sharp.
On the other hand, $R_{m^{2}}(U,L)$ decreases to 0 as $L \to \infty$
in the disordered phase.
At the critical point, the correlation ratio becomes 
volume independent;
it is expected to cross at a universal value for different $L$,
by which feature we can determine the critical point 
in a sensitive way~\cite{Kaul_PRL2015,Pujari_PRL2016}.
However, if there are non-negligible corrections to the scaling, 
the crossing points of $R_{m^{2}}(U,L)$ drift with increasing $L$.
We clearly observe this drift as shown in Fig.~\ref{fig:crossing_points},
which is in contrast with the recent study of SLAC fermions~\cite{Lang_PRL2019}.
In the presence of the corrections,
the crossing-point analysis~\cite{Shao_Science2016} 
serves as a simple and reliable way to find the critical point,
which has been developed mainly in 
quantum spin systems~\cite{Wang_PRB2006,Wenzel_PRB2009,Jiang_2009,Shao_Science2016,Ma_PRL2018,Ran_PRB2019}
and has also recently been applied to 
fermionic models~\cite{ParisenToldin_PRB2015,Hesselmann_PRB2016,Liu_Natcomm2019,Liu_PRB2020,Li_PRB2020,Xu_arXiv2019,Sato_arXiv2020}.

\begin{figure}[tb]
 \centering
 \includegraphics[width=1.00\linewidth]{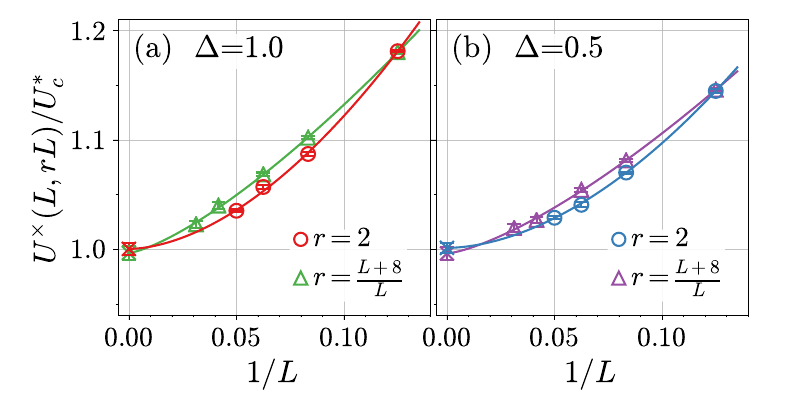}
 \caption{\label{fig:Ux-L}%
 Crossing-point analysis of the correlation ratio.
 $U^{\times}(L, rL)$ are obtained by interpolating 
 data points of $R_{m^{2}}(U,L)$ with second-order polynomial functions.
 For ease of comparison, values of $U^{\times}(L, rL)$ are
 normalized by $U^{\ast}_{\mathrm{c}}=7.63$ and 5.49
 for (a) $\Delta=1$ and (b) $\Delta=0.5$, respectively.
 Ratios of two system sizes are $r=2$ (circles) 
 and $r=(L+8)/L$ (triangles).
 Results of the fit with
 $U^{\times}(L, rL) = U_{\mathrm{c}} + d L^{-(\omega + 1/\nu)}$
 are
 (a) $U_{\mathrm{c}}=7.63(4)$ and $\omega+1/\nu=1.8(1)$ for $r=2$
 and $U_{\mathrm{c}}=7.61(5)$ and $\omega+1/\nu=1.4(1)$ for $r=(L+8)/L$,
 and
 (b) $U_{\mathrm{c}}=5.49(3)$ and $\omega+1/\nu=1.8(1)$  for $r=2$
 and $U_{\mathrm{c}}=5.47(3)$ and $\omega+1/\nu=1.4(1)$  for $r=(L+8)/L$.
 Numbers in parentheses denote errors in the last digits.
 Note that $U_{\mathrm{c}}$s extrapolated in the thermodynamic limit are 
 indicated by crosses at $1/L=0$.
 }
\end{figure}

A crossing point denoted by $U^{\times}(L, rL)$ is defined as 
a value of $U$ at which two curves of $R_{m^{2}}(U,L)$ and $R_{m^{2}}(U,rL)$ cross,
which we compute by polynomial interpolation.
Based on the phenomenological renormalization argument,
the crossing points are extrapolated to the critical point 
following a simple power law~\cite{Shao_Science2016}:
\begin{equation}
 U^{\times}(L, rL) = U_{\mathrm{c}} + c L^{-(\omega + 1/\nu)},
  \label{eq:Ux}
\end{equation}
where $c$ is a constant, $\omega$ is an exponent of the 
leading correction term, 
and $\nu$ is the correlation-length exponent.
We choose the ratios of the two system sizes as $r=2$ and $r=(L+8)/L$,
each of which covers all the lattice sizes $L$ (= 8, 12, 16, 20, 24, 32, 40).
The results show that
both series for each $\Delta$ are extrapolated to the same $U_{\mathrm{c}}$ 
within the error bars;
$U_{\mathrm{c}}=7.63(4)$ and 7.61(5) for $\Delta=1$ [Fig.~\ref{fig:Ux-L}(a)], 
and
$U_{\mathrm{c}}=5.49(3)$ and 5.47(3) for $\Delta=0.5$ [Fig.~\ref{fig:Ux-L}(b)].
The fitted exponents in Eq.~(\ref{eq:Ux}), i.e., $\omega+1/\nu$
fall in almost the same value for each $r$ irrespectively of $\Delta$;
$\omega+1/\nu=1.8(1)$ for $r=2$, 
and
$\omega+1/\nu=1.4(1)$ for $r=(L+8)/L$.
If the phase transitions for $\Delta=1$ and $0.5$ belong to the same universality class, 
both $\nu$ and $\omega$ should be universal, resulting in the same value of $\omega+1/\nu$.
This is indeed the case in our result for each $r$. 
The difference in $\omega+1/\nu$ between $r=2$ and $(L+8)/L$ 
is speculated to be due to large corrections to the scaling.
Even with the large corrections, 
Eq.~(\ref{eq:Ux}) can be exploited by considering
the exponent $\omega$ as an effective one
which implicitly includes effects of higher-order corrections~\cite{Shao_Science2016,Ran_PRB2019}.
Our result may imply that $\nu$ is universal with the effective $\omega$ 
which is dependent on the fitting condition, such as the choice of $r$.

\begin{figure}[tb]
 \centering
 \includegraphics[width=1.0\linewidth]{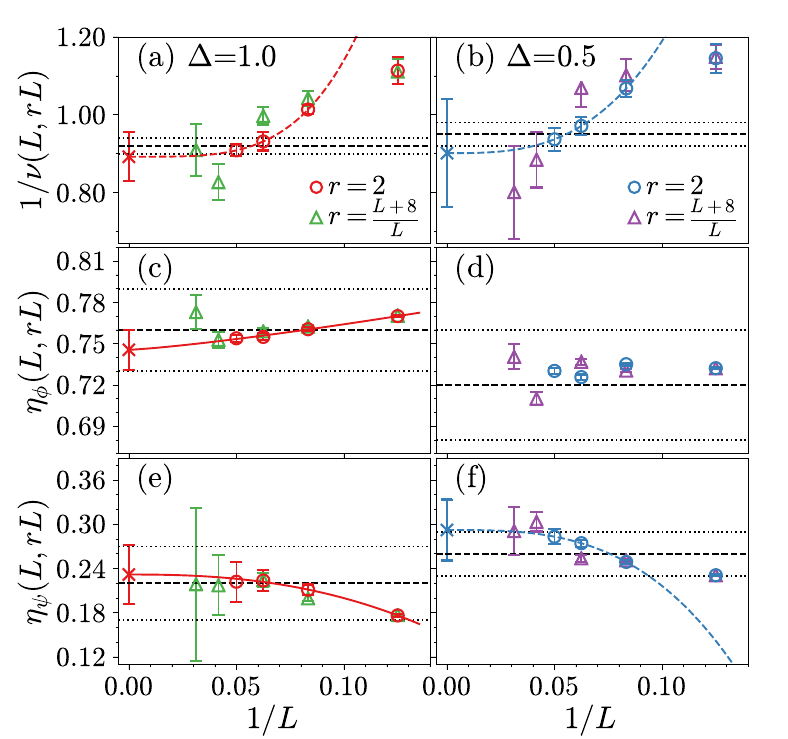}
 \caption{\label{fig:exponents-L}%
 System size dependence of 
 (a), (b)
 the inverse correlation-length exponent $1/\nu$,
 (c), (d)
 the anomalous dimension for the order parameter $\eta_{\phi}$
 , and
 (e), (f)
 the anomalous dimension for the fermionic field $\eta_{\psi}$.
 Left (right) panels show results for $\Delta=1$ (0.5).
 Ratios of two system sizes are $r=2$ (circles) and $r=(L+8)/L$ (triangles).
 Solid and dashed lines are fits to data of $r=2$ with and without $L=8$, respectively,
 from which the exponents are estimated as
 (a) $\nu         = 1.12(8)$, 
 (b) $\nu         = 1.11(17)$, 
 (c) $\eta_{\phi} = 0.75(2)$, 
 (e) $\eta_{\psi} = 0.23(4)$, and
 (f) $\eta_{\psi} = 0.29(4)$,
 indicated by crosses at $1/L=0$.
 Here, numbers in parentheses denote errors in the last digits.
 In the other cases, fitted lines are not shown, 
 because the fits are unstable.
 Horizontal dashed and dotted lines represent 
 values of the exponents  and their error bars 
 estimated from slopes of log-log plots of
 $m^{2}(U_{\mathrm{c}},L)$, $Z(U_{\mathrm{c}},L)$, and $s(U_{\mathrm{c}}, L)$
 in Figs.~\ref{fig:eta-Ux_D010},~\ref{fig:eta-Ux_D005}, and ~\ref{fig:s-L}.
 }
\end{figure}

The correlation-length exponent $\nu$ itself can be estimated 
by the crossing-point analysis~\cite{Shao_Science2016}.
Since we have used the polynomial interpolation of $R_{m^{2}}(U,L)$ 
to find $U^{\times}(L, rL)$, it is straightforward to compute its slope,
$s(U,L)=\frac{d R_{m^{2}}(U,L)}{dU}$.
The size-dependent inverse correlation-length exponent is calculated
from ratio of the two slopes at the crossing point,
\begin{equation}
 \frac{1}{\nu(L, rL)} = \frac{1}{\ln(r)}
  \ln \left\{
  \frac{s(U^{\times}, rL)}{s(U^{\times}, L)}
  \right\},
  \label{eq:nu-L}
\end{equation}
where $U^{\times}$ stands for a shortened form of $U^{\times}(L, rL)$.
This quantity scales to the correct exponent at the rate $L^{-\omega}$,
\begin{equation}
  \frac{1}{\nu(L, rL)} = \frac{1}{\nu} + d L^{-\omega},
\end{equation}
where $d$ is a constant and $\omega$ is the correction exponent.
Since the evaluation of the slope is rather sensitive and affected by
the small number of the data points of $R_{m^{2}}(U,L)$,
the fit to the data turns out to be difficult.
Nonetheless, the fit without $L=8$ for $r=2$ yields
consistent values of $\nu=1.12(8)$ and 1.11(17) 
for $\Delta=1$ and 0.5, respectively,
as shown in Figs.~\ref{fig:exponents-L}(a) and \ref{fig:exponents-L}(b).
The correction exponent $\omega$ is expected to be the same as that 
in Eq.~(\ref{eq:Ux}), especially since basically the same quantity, 
$R_{m^{2}}(U,L)$, is examined here.
However, the resultant $\omega$ 
in Figs.~\ref{fig:exponents-L}(a) and \ref{fig:exponents-L}(b) are larger than 
those expected from $\omega+1/\nu=1.4 - 1.8$ in Fig.~\ref{fig:Ux-L}
and $\nu \simeq 1.1$.
The analysis based on Eq.~(\ref{eq:Ux}) is more straightforward, and 
thereby we expect that it is more reliable than 
the analysis based on Eq.~(\ref{eq:nu-L}).
The exponent $\omega$ in the latter case
should be considered as an independent fitting parameter
to estimate $\nu$.

\subsection{scaling at the critical point}\label{subsec:poorman}

In the GN scenario,
the phase transition, namely spontaneous symmetry breaking,
accompanies with opening a charge gap~\cite{Gross_PRD1974,Ryu_PRB2009}.
Thus, it is characterized by two quantities,
the staggered magnetization $m^{2}(U,L)$ 
and the quasiparticle weight $Z(U,L)$.
In the QMC simulations, $m^{2}(U,L)$ is calculated from 
the spin structure factor in Eq.~(\ref{eq:Sk}) as
\begin{equation}
m^{2}(U,L)=S(\bm{K})/L^{2} \label{eq:m2}. 
\end{equation}
$Z(U,L)$ is also estimated as follows~\cite{Seki_PRB2019,Otsuka_PRB2018}:
\begin{equation}
 Z(U,L) = \frac{D_{\sigma}(U, L)}{D_{\sigma}(0, L)}, \label{eq:zul}
\end{equation}
where 
$D_{\sigma}(U, L) = \frac{1}{L^{2}}\sum_{i} 
\langle c_{j \sigma}^{\dagger} c_{i \sigma} \rangle$ 
is the equal-time single-particle Green's function
at the maximum distance, 
$\bm{r}_{j}- \bm{r}_{i}=(L/2, L/2)$.
At the critical point, these two quantities scale as
\begin{equation}
 m^{2}(U_{\mathrm{c}},L) \propto L^{-1-\eta_{\phi}} \left( 1 + g L^{-\omega} \right),
 \label{eq:m2-L} 
\end{equation}
and
\begin{equation}
 Z(U_{\mathrm{c}},L)     \propto L^{-\eta_{\psi}}   \left( 1 + h L^{-\omega} \right),
 \label{eq:Z-L} 
\end{equation}
where $\eta_{\phi}$ ($\eta_{\psi}$) is the anomalous dimension
for the order parameter (fermionic field), 
and the effective correction terms to the first order 
are taken into account with $g$ and $h$ being constants.
The exponents $\omega$ are taken into account as independent fitting parameters.
Similar to Eq.(\ref{eq:nu-L}),
the size-dependent exponents can be defined 
by taking the two system sizes $L$ and $rL$ as
\begin{equation}
 \eta_{\phi}(L, rL) =
  \frac{1}{\ln(r)}
  \ln
  \left\{ \frac{m^{2}(U_{\mathrm{c}}, L)} {m^{2}(U_{\mathrm{c}}, rL)} \right\} - 1
 \label{eq:def-eta_phi-L} 
\end{equation}
and
\begin{equation}
 \eta_{\psi}(L, rL) =
  \frac{1}{\ln(r)}
 \ln \left\{ \frac{Z(U_{\mathrm{c}}, L)} {Z(U_{\mathrm{c}}, rL)} \right\}.
 \label{eq:def-eta_psi-L} 
\end{equation}
From these quantities, the exponents are extrapolated as
\begin{equation}
 \eta_{\phi}(L, rL) = \eta_{\phi} + g^{\prime} L^{-\omega}
 \label{eq:eta_phi-L}
\end{equation}
and
\begin{equation}
 \eta_{\psi}(L, rL) = \eta_{\psi} + h^{\prime} L^{-\omega},
 \label{eq:eta_psi-L}
\end{equation}
where $g^{\prime}$ and $h^{\prime}$ are constants.
The fits of the data according to these forms
are again rather difficult,
as shown in Figs.~\ref{fig:exponents-L}(c)-\ref{fig:exponents-L}(f),
some of which however yield the estimations as
$\eta_{\phi}=0.75(2)$ for $\Delta=1$   and $r=2$,
$\eta_{\psi}=0.23(4)$ for $\Delta=1$   and $r=2$, and
$\eta_{\psi}=0.29(4)$ for $\Delta=0.5$ and $r=2$.
Although the estimate of $\eta_{\phi}$ for $\Delta=0.5$ is not available
because of limitation in the quality of data, 
it seems consistent with that for $\Delta=1$.

\begin{figure}[tb]
 \centering
 \includegraphics[width=1.0\linewidth]{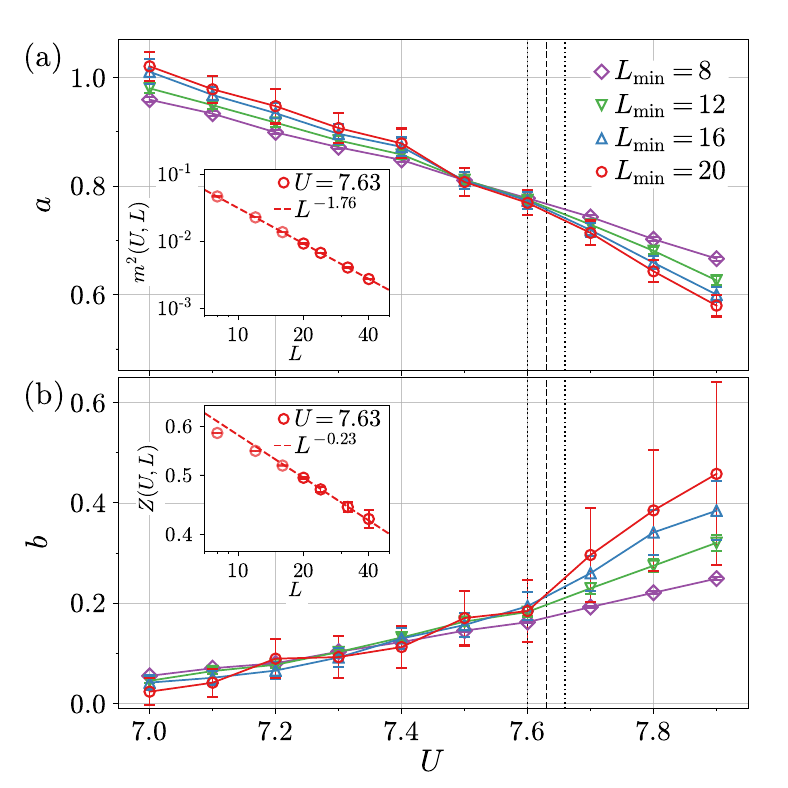}
  \caption{\label{fig:eta-Ux_D010}%
  $U$ dependence of fitted exponents for $\Delta=1$;
  (a) staggered magnetization $m^{2}(U,L)$ is fit to $A L^{-1-a}$ and
  (b) quasiparticle weight $Z(U,L)$        is fit to $B L^{-b}$.
  Fitting range is between $L_{\mathrm{min}}$ and $L_{\mathrm{max}}\, (=40)$.
  Vertical dashed and dotted lines indicate 
  the critical point and the error bar, 
  $U^{\ast}_{\mathrm{c}}=7.63(4)$,
  estimated from the crossing-point analysis of the correlation ratio
  in Fig.~\ref{fig:Ux-L}(a).
  Insets of panels (a) and (b) are log-log plots of 
  $m^{2}(U^{\ast}_{\mathrm{c}},L)$ with $\eta_{\phi}=0.76(3)$
  and 
  $Z(U^{\ast}_{\mathrm{c}},L)$ with $\eta_{\psi}=0.22(5)$, respectively.
  Here, numbers in parentheses denote errors in the last digits.
 }
\end{figure}

\begin{figure}[tb]
 \centering
 \includegraphics[width=1.0\linewidth]{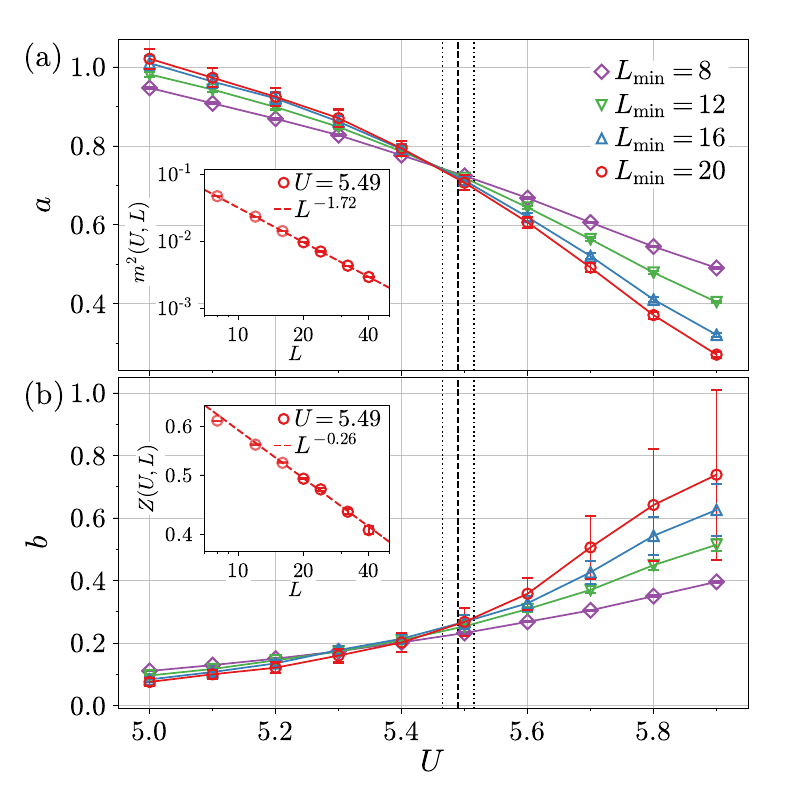}
 \caption{\label{fig:eta-Ux_D005}%
 $U$ dependence of fitted exponents for $\Delta=0.5$;
 (a) staggered magnetization $m^{2}(U,L)$ is fit to $A L^{-1-a}$ and
 (b) quasiparticle weight $Z(U,L)$        is fit to $B L^{-b}$.
 Fitting range is between $L_{\mathrm{min}}$ and $L_{\mathrm{max}}\, (=40)$.
 Vertical dashed and dotted lines indicate 
 the critical point and the error bar, 
 $U^{\ast}_{\mathrm{c}}=5.49(3)$,
 estimated from the crossing-point analysis of the correlation ratio
 in Fig.~\ref{fig:Ux-L}(b).
 Insets of panels (a) and (b) are log-log plots of 
 $m^{2}(U^{\ast}_{\mathrm{c}},L)$ with $\eta_{\phi}=0.72(4)$
 and 
 $Z(U^{\ast}_{\mathrm{c}},L)$     with $\eta_{\psi}=0.26(3)$, respectively.
 Here, numbers in parentheses denote errors in the last digits.
 }
\end{figure}

We also estimate the exponents $\eta_{\phi}$ and $\eta_{\psi}$
by a more naive method 
without assuming a specific form of the correction term.
Although $m^{2}(U, L)$ is expected to decay as $L^{-1-\eta_{\phi}}$
only at $U=U_{\mathrm{c}}$ and for large $L$, 
here we try to fit the data points of $m^{2}(U, L)$  
to a function $A L^{-1-a}$ with two, $A$ and $a$, fitting parameters,
regardless of whether $U$ is close to $U_{\mathrm{c}}$. 
At each $U$, the range of the fit is chosen as $L \in [L_{\mathrm{min}}, L_{\mathrm{max}}(=40)]$
and we examine how the fitted exponent $a$ changes
with increasing $L_{\mathrm{min}}$.
Since $m^{2}(U=0, L)$ decays as $L^{-2}$,
$a$ is expected to approach 1 for $U<U_{\mathrm{c}}$,
whereas it should converge to $\eta_{\phi}$ at $U=U_{\mathrm{c}}$.
As shown in Figs.~\ref{fig:eta-Ux_D010}(a) and \ref{fig:eta-Ux_D005}(a),
the asymptotic behavior of $a$ indeed changes at $U \simeq U^{\ast}_{\mathrm{c}}$ 
that is estimated from the crossing-point analysis in Fig.~\ref{fig:Ux-L}.
At these critical points, 
we estimate $\eta_{\phi}$ from the slope of the log-log plots of $m^{2}(U=U^{\ast}_{\mathrm{c}}, L)$ 
for $L \ge L_{\mathrm{min}}=20$ 
[see insets of Figs.~\ref{fig:eta-Ux_D010}(a) and \ref{fig:eta-Ux_D005}(a)]
as $\eta_{\phi}=0.76(3)$ and $0.72(4)$ for $\Delta=1$ and $0.5$,
respectively.
These values are consistent with the estimations from 
Eqs.~(\ref{eq:eta_phi-L}) and (\ref{eq:eta_psi-L}) 
[also see Figs.~\ref{fig:exponents-L}(c) and \ref{fig:exponents-L}(d)] 
and imply that this exponent is the same, independently of $\Delta$.

We apply the same analysis to the quasiparticle weight; 
$Z(U,L)$ is fit to the function $BL^{-b}$.
Note that, in this case, since for $U<U_{\mathrm{c}}$ 
the equal-time single-particle Green's function 
$D_{\sigma}(U,L)$ in Eq.~(\ref{eq:zul}) decreases as $r^{-2}$ 
with $r$ being the distance and
for $U>U_{\mathrm{c}}$ it decays exponentially~\cite{Seki_PRB2019},
the fitted exponent $b$ shows an abrupt change at $U \simeq U^{\ast}_{\mathrm{c}}$,
as shown in Figs.~\ref{fig:eta-Ux_D010}(b) and \ref{fig:eta-Ux_D005}(b).
This is naturally expected from the GN scenario that describes
SM and a gapped ordered phase separated by a single phase transition,
and at the same time provides further evidence for the accuracy of 
estimates of the critical point.
The critical exponents estimated from the log-log plots at $U=U^{\ast}_{\mathrm{c}}$,
as shown in the insets of Figs.~\ref{fig:eta-Ux_D010}(b) and \ref{fig:eta-Ux_D005}(b),
again fall into the same value within the error bars;
$\eta_{\psi}=0.23(5)$ and $0.26(3)$ for $\Delta=1$ and $0.5$,
respectively.

\begin{figure}[tb]
 \centering
 \includegraphics[width=0.625\linewidth]{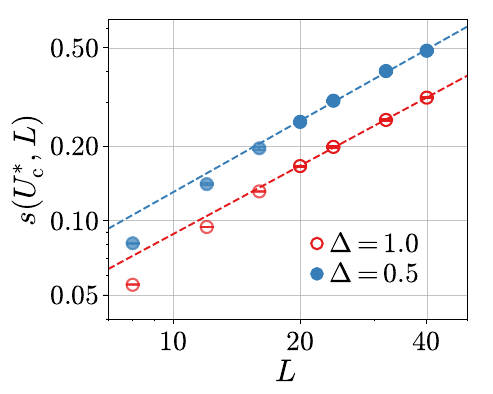}
 \caption{\label{fig:s-L}%
 Log-log fits of slopes of correlation ratio at the estimated critical point $U^{\ast}_{\mathrm{c}}$.
 Open (solid) circles represent results for $\Delta=1$ (0.5)  with $U^{\ast}_{\mathrm{c}}=7.63$ (5.49).
 Dashed lines are fits to data for $L \ge 20$.
 Estimations of $\nu$ are 1.09(2) and 1.04(3) for $\Delta=1$ and 0.5, 
 respectively.
 Here, numbers in parentheses denote errors in the last digits.
 }
\end{figure}

Since it turns out that our $U^{\ast}_{\mathrm{c}}$ is a good estimation of $U_{\mathrm{c}}$ 
and the conventional log-log fits given above work well,
we may as well evaluate $1/\nu$ from the log-log fit of $s(U,L)$, 
which is expected to behave at the critical point for large $L$ 
as follows~\cite{Wenzel_PRB2009}:
\begin{equation}
 s(U_{\mathrm{c}},L) \sim L^{1/\nu}.
\label{eq:s-L}
\end{equation}
From the fit in Fig.~\ref{fig:s-L},
the exponent is obtained as 
$\nu=1.09(2)$ and 1.04(3) for $\Delta=1$ and 0.5, respectively,
being consistent with the crossing-point analysis of $\nu(L,rL)$ shown in 
Figs.~\ref{fig:exponents-L}(a) and \ref{fig:exponents-L}(b).

\subsection{data collapse}\label{subsec:data_collapse}

Finally, the critical points and exponents are also assessed by 
collapsing data on the basis of the finite-size scaling ansatz,
\begin{align}
 R_{m^{2}}(U,L)   &=                    f_{R}( u L^{1/\nu}),\\
 m^{2}(U,L) &= L^{-1-\eta_{\phi}} f_{m}( u L^{1/\nu}),
\end{align}
and
\begin{equation}
 Z(U,L)     = L^{  -\eta_{\psi}} f_{Z}( u L^{1/\nu}),
\end{equation}
where $u=(U-U_{\mathrm{c}})/U_{\mathrm{c}}$ represents normalized distance from the critical point
and $f_{\alpha}$ ($\alpha=R, m, Z$) the scaling functions.
The correction terms are not explicitly included in the scaling form.
Instead, as described in the second half of Sec.~\ref{subsec:poorman}, 
we perform data collapsing for $L$ between $L_{\mathrm{min}}$ and $L_{\mathrm{max}}$
and analyze trends of fit results with increasing $L_{\mathrm{min}}$~\cite{Otsuka_PRX2016,Otsuka_PRB2018}.

\begin{figure}[tb]
 \centering
 \includegraphics[width=1.0\linewidth]{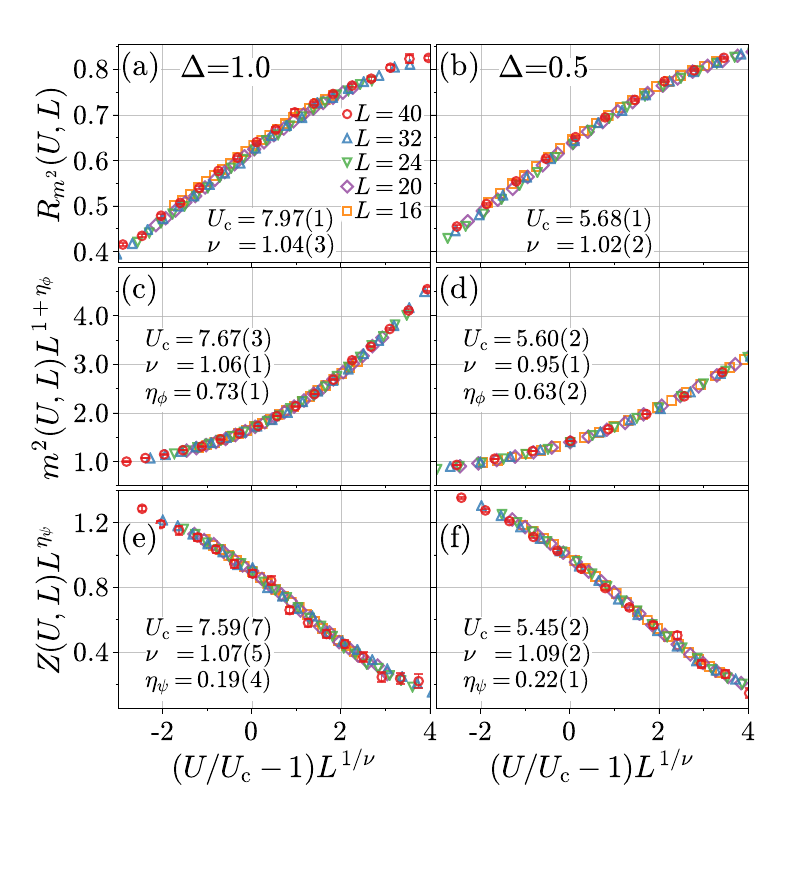}
 \caption{\label{fig:collapse}%
 Data-collapse fits of 
 (a), (b)
 correlation ratio,
 (c), (d)
 staggered magnetization,
 and
 (e), (f)
 quasiparticle weight.
 For each observable,
 left and right figures show results of $\Delta=1$ and 0.5
 with the same scale.
 Estimated critical points and exponents are indicated 
 in each figure.
 The number in parentheses indicates the statistical error, 
 corresponding to the last digit of the value,
 which is estimated by the resampling technique~\cite{Otsuka_PRX2016}.
 }
\end{figure}

We employ the Bayesian method to tightly collapse the data
without relying on using a specific polynomial function~\cite{Harada_PRE2011}.
Figure~\ref{fig:collapse} shows typical examples of the data collapses for $L_{\mathrm{min}}=16$.
From each of the collapse fits, 
the sets of $U_{\mathrm{c}}$ and $\nu$ are independently estimated.  
In addition, $\eta_{\phi}$ and $\eta_{\psi}$ are also determined from the data collapse fits of 
$m^{2}(U,L)$ and $Z(U,L)$, respectively.
Before analyzing further details, it is worth noticing that, 
for $\Delta=1$ and 0.5, 
the scaling functions for each observable seem similar and apparently  
superpose to each other by considering an appropriate
rescaling with nonuniversal metric factors, which indicates
the existence of the universality class~\cite{Privman_PRB1984}.

\begin{figure}[tb]
 \centering
 \includegraphics[width=0.75\linewidth]{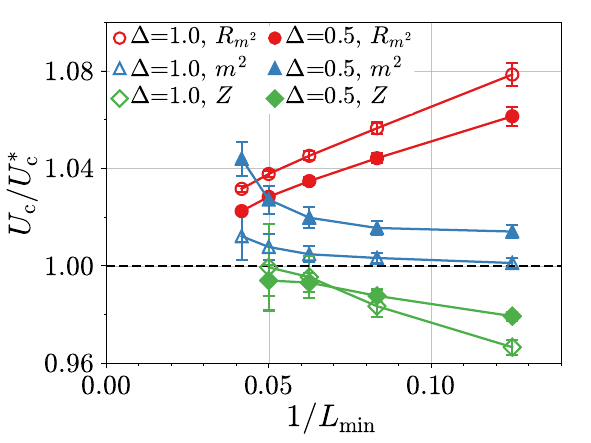}
 \caption{\label{fig:vsLmin_Uc}%
 $1/L_{\mathrm{min}}$ dependencies of the critical points $U_{\mathrm{c}}$
 estimated from data-collapse fits of
 correlation ratio       (circles),
 staggered magnetization (triangles),
 and
 quasiparticle weight     (diamonds).
 For ease of comparison, 
 values of $U_{\mathrm{c}}$ are
 normalized by $U^{\ast}_{\mathrm{c}}=7.63$ and 5.49
 for $\Delta=1$ (open   symbols) 
 and $\Delta=0.5$ (solid symbols), respectively.
 }
\end{figure}

As shown in Fig.~\ref{fig:vsLmin_Uc},
the estimations of $U_{\mathrm{c}}$ obtained by collapsing $R_{m^{2}}(U,L)$ comparatively
depend on $L_{\mathrm{min}}$; however their dependency seems to be controllable,
evenly approaching $U^{\ast}_{\mathrm{c}}$ that are determined by the crossing-point 
analysis in Fig.~\ref{fig:Ux-L}.
The other estimations of $U_{\mathrm{c}}$ show the convergence to $U^{\ast}_{\mathrm{c}}$ already 
for $L_{\mathrm{min}}=16$, except for the case of $m^{2}(U,L)$ for $\Delta=0.5$.

\begin{figure}[tb]
 \centering
 \includegraphics[width=0.75\linewidth]{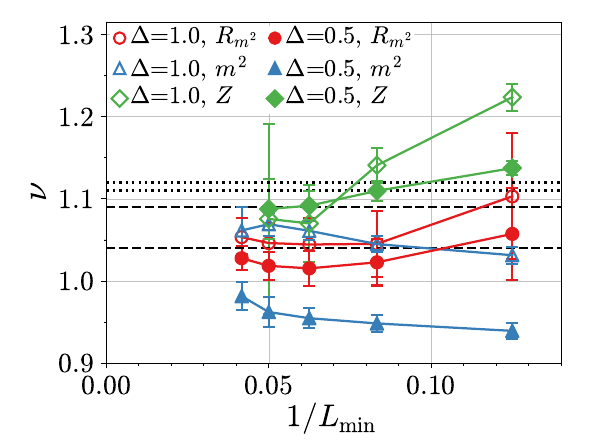}
 \caption{\label{fig:vsLmin_nu}%
 $1/L_{\mathrm{min}}$ dependencies of the correlation-length exponent $\nu$
 estimated from data-collapse fits of
 correlation ratio       (circles),
 staggered magnetization (triangles),
 and
 quasiparticle weight     (diamonds).
 Open (solid) symbols represent the results for $\Delta=1$ (0.5).
 Dotted and dashed lines represent the estimations obtained 
 from the crossing-point analysis 
 in Figs.~\ref{fig:exponents-L}(a) and \ref{fig:exponents-L}(b),  and 
 from the scaling at the critical point 
 in Fig.~\ref{fig:s-L}, respectively.
 }
\end{figure}

Similar tendencies are observed in $\nu$ (see Fig.~\ref{fig:vsLmin_nu}).
For $L_{\mathrm{min}} \ge 16$, the values of $\nu$ fall almost between 1.0 and 1.1.
Although it is not trivial to judge which estimation is most reliable,
the candidate would be that obtained from $R_{m^{2}}(U,L)$ for $\Delta=1$,
i.e., $\nu \simeq 1.05$,
because the collapse fit of $R_{m^{2}}(U,L)$ includes the smaller number of the 
fitting parameters than the other observables,
and the system with the isotropic Dirac cones for $\Delta=1$ may have 
fewer correction effects.

\begin{figure}[tb]
 \centering
 \includegraphics[width=0.75\linewidth]{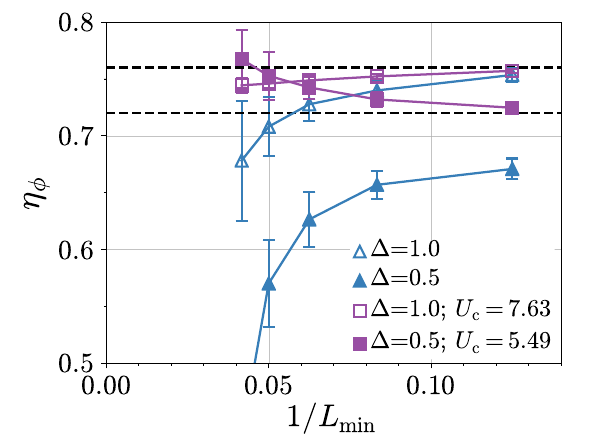}
 \caption{\label{fig:vsLmin_eta_phi}%
 $1/L_{\mathrm{min}}$ dependencies of the anomalous dimension of the order parameter $\eta_{\phi}$
 estimated from data-collapse fits of 
 staggered magnetization.
 Open (solid) triangles represent the results for $\Delta=1$ (0.5).
 Results of the data-collapse fits with fixed $U_{\mathrm{c}}=U^{\ast}_{\mathrm{c}}=7.63$ (5.49)
 for $\Delta=1$ (0.5) are shown by open (solid) squares.
 Dashed lines represent the estimations obtained from 
 the scaling at the critical points in 
 Figs.~\ref{fig:eta-Ux_D010}(a) and \ref{fig:eta-Ux_D005}(a).
 }
\end{figure}

\begin{figure}[tb]
 \centering
 \includegraphics[width=0.75\linewidth]{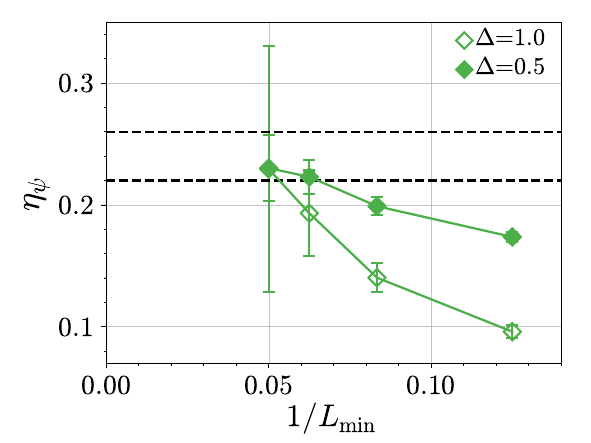}
 \caption{\label{fig:vsLmin_eta_psi}%
 $1/L_{\mathrm{min}}$ dependencies of the anomalous dimension of the fermionic field $\eta_{\psi}$
 estimated from data-collapse fits of 
 quasiparticle weight. 
 Open (solid) diamonds represent the results for $\Delta=1$ (0.5).
 Dashed lines represent the estimations obtained from 
 the scaling at the critical points in 
 Figs.~\ref{fig:eta-Ux_D010}(b) and \ref{fig:eta-Ux_D005}(b).
 }
\end{figure}

The deviation of the exponents estimated from $m^{2}(U,L)$ for $\Delta=0.5$
is further noticeable in $\eta_{\phi}$, as shown in Fig.~\ref{fig:vsLmin_eta_phi}.
To check which estimated values of $\eta_{\phi}$ are closer to the exact one, 
we also perform the data-collapse fits of $m^{2}(U,L)$ with $U_{\mathrm{c}}$ fixed at $U^{\ast}_{\mathrm{c}}$, 
which is the most accurately obtained quantity in this study 
owing to the well-designed crossing-point analysis, whereas $U_{\mathrm{c}}$ estimated 
from the data-collapse fits of $m^{2}(U,L)$ sizably deviates  from $U^{\ast}_{\mathrm{c}}$ for $\Delta=0.5$ 
(see Fig.~\ref{fig:vsLmin_Uc}).
As shown in Fig.~\ref{fig:vsLmin_eta_phi}, the exponent is converged to 
$\eta_{\phi} \simeq 0.75$ for both $\Delta=1$ and 0.5. 
This value is closer to the estimation of the isotropic case ($\Delta=1$) with 
unfixed $U_{\mathrm{c}}$ and also to the results obtained from the scaling at the critical points 
shown in Figs.~\ref{fig:eta-Ux_D010}(a) and ~\ref{fig:eta-Ux_D005}(a).
Finally, Fig.~\ref{fig:vsLmin_eta_psi} presents the results of $\eta_{\psi}$ 
estimated from the data-collapse fits of $Z(U,L)$,
in which the values of $\eta_{\psi}$ for $\Delta=1$ and 0.5 are confirmed 
to be coincident with each other.

 \section{Discussion and Conclusions}\label{sec:conclusion}

\begin{table}[tb]
 \caption{\label{tbl:exponents}%
 Results of the critical exponents.
 Estimations obtained in this work are summarized in the upper group.
 For comparison, previous QMC results for other lattice models and 
 recent analytical results for the Gross-Neveu model
 are also listed in the middle and lower groups, respectively.
 }
 \begin{ruledtabular}
  \begin{tabular}{l l l l l}
   Method                      &        & $\nu$        &$\eta_{\phi}$  &$\eta_{\psi}$ \\
   \hline
   AFQMC (this work)           &$\Delta$&              &               &          \\
   \cline{1-2}
   $1/L \to 0$ of $L$ and $2L$ & 1      & 1.12(8)      & 0.75(2)       & 0.23(4)  \\
                               & 0.5    & 1.11(17)     & n.a.          & 0.29(4)  \\
   log-log fit   at $U^{\ast}_{\mathrm{c}}$     & 1      & 1.09(2)      & 0.76(3)       & 0.22(5)  \\
                               & 0.5    & 1.04(3)      & 0.72(4)       & 0.26(3)  \\
   data-collapse$^{a}$ of $R_{m^{2}}$& 1      & 1.05(2)      & ---           & ---      \\
                               & 0.5    & 1.02(2)      & ---           & ---      \\
   data-collapse$^{a}$ of $m^{2}$& 1      & 1.07(1)      & 0.71(3)       & ---      \\
                               & 1      &              & 0.75(1)$^{b}$ & ---      \\
                               & 0.5    & 0.96(2)      & 0.57(4)       & ---      \\
                               & 0.5    &              & 0.75(2)$^{b}$ & ---      \\
   data-collapse$^{a}$ of $Z$  & 1.0    & 1.08(12)     & ---           & 0.23(10) \\
                               & 0.5    & 1.09(4)      & ---           & 0.23(3)  \\
   \\
   \multicolumn{2}{l}{AFQMC$^{c}$~\cite{Otsuka_PRX2016}} &  
   1.02(1)       &
   0.49(4)       &  
   0.20(2)       \\
   \multicolumn{2}{l}{}                                 & 
                 &
   0.65(3)$^{e}$ & 
			     \\
   \multicolumn{2}{l}{AFQMC$^{d}$~\cite{Otsuka_PRX2016}}  &
   1.02(1)       & 
   0.45(6)       & 
   0.23(2)       \\
   \multicolumn{2}{l}{}                                 & 
                 &
   0.64(6)$^{e}$ & 
                 \\
   \multicolumn{2}{l}{AFQMC$^{c}$~\cite{ParisenToldin_PRB2015}} &
		   0.84(4) & 0.70(15) &             \\  
   \multicolumn{2}{l}{AFQMC$^{f}$~\cite{Liu_Natcomm2019}} &
		   0.88(7) & 0.79(5)  &             \\  
   \multicolumn{2}{l}{AFQMC$^{g}$~\cite{Xu_arXiv2020}} &
		   0.99(8) & 0.55(2)  &             \\  
   \multicolumn{2}{l}{HMC$^{c}$~\cite{Buividovich_PRB2018}} &
		   1.162 & 0.872(44)  &             \\
   \\
   \multicolumn{2}{l}{$4-\epsilon$, 4th order~\cite{Zerf_PRD2017}} & 
		   1.2352 & 0.9563 &  0.1560  \\
   \multicolumn{2}{l}{FRG~\cite{Knorr_PRB2018}} &
		   1.26   & 1.032  & 0.071(2) \\
   \multicolumn{2}{l}{FRG~\cite{Janssen_PRB2014}} &
            1.314 & 1.012 & 0.083  \\
   \multicolumn{2}{l}{Large $N$~\cite{Gracey_PRD2018}} &
           1.1823 & 1.1849 & 0.1051 \\
  \end{tabular}
  {\raggedright
  $^{a}$ $L_{\mathrm{min}}=20$.\\
  $^{b}$ $U_{\mathrm{c}}$ is fixed at $U^{\ast}_{\mathrm{c}}=7.63$ (5.49) for $\Delta=1$ (0.5).\\
  $^{c}$ SM-AF transition in the honeycomb lattice model.\\
  $^{d}$ SM-AF transition in the $\pi$-flux model.\\
  $^{e}$ Estimated by collapsing $m^{2}$ without the correction term.\\
  $^{f}$ SM to quantum spin-Hall transition on the honeycomb lattice.\\
  $^{g}$ $d$-wave SC to coexistence phase with gapped $d$-wave SC and AF order.
  \par}
 \end{ruledtabular}
\end{table}

Table~\ref{tbl:exponents} summarizes 
the critical exponents of our model estimated by the various methods. 
These exponents are also 
compared with those obtained by QMC calculations on other lattice
models~\cite{ParisenToldin_PRB2015,Otsuka_PRX2016,Buividovich_PRB2018,Liu_Natcomm2019}
and by recent analytical studies
on the GN model~\cite{Zerf_PRD2017,Knorr_PRB2018,Janssen_PRB2014,Gracey_PRD2018},
all for the $N=8$ chiral-Heisenberg universality class.

Within the results of this work,
the agreement of each exponent is mostly confirmed irrespectively of $\Delta$.
To be precise, the exception is found in the collapse fit of $m^{2}(U,L)$
for $\Delta=0.5$. However, the difference in $\nu$ is marginally
within two standard deviations, and the discrepancy in $\eta_{\phi}$ is
resolved by fixing $U_{\mathrm{c}}=U^{\ast}_{\mathrm{c}}$.
Therefore, our results represent accurate estimations 
of the critical exponents describing the chiral-Heisenberg universality class.
It is also stressed that the exponents are shared with the system with 
the anisotropic Dirac cones with $\Delta=0.5$.
In the previous studies, e.g., Ref.~[\onlinecite{Otsuka_PRX2016}],
it was shown that the isotropic Dirac Fermi velocity $v_{\mathrm{F}}^{0}$, 
which is different between the honeycomb lattice model and the $\pi$-flux model,
does not affect the criticality.
In addition, the present results indicate that the anisotropic 
velocity around the Dirac point is also irrelevant 
for the nature of the fixed point~\cite{Sachdev_Book1999,Vafek_PRL2002,Sharma_arXiv2012,Wang_PRB2014,Roy_JHEP2015}.

Our results are comparable to the numerical results of the different lattice
models~\cite{ParisenToldin_PRB2015,Otsuka_PRX2016,Buividovich_PRB2018,Liu_Natcomm2019}.
The estimations of $\nu$ and $\eta_{\psi}$ are almost identical to 
the previous results that were reported by some of the present authors~\cite{Otsuka_PRX2016}, 
while noticeable discrepancy is observed in $\eta_{\phi}$.
We here point out that 
the treatment of the correction term in the collapse fit of the staggered
magnetization was presumably inaccurate, and instead the results of the collapse 
fits without the correction term for large $L_{\mathrm{min}}$, 
shown in Table~\ref{tbl:exponents} 
as well as in Table~II of Ref.~[\onlinecite{Otsuka_PRX2016}], 
are more reliable.
More detailed reassessment on this point is left for future work.
The estimated $\nu$ and $\eta_{\phi}$ of this work are also consistent with
the recent AFQMC study on the Kane-Mele-like model showing a phase transition 
between SM and a quantum spin Hall insulator~\cite{Liu_Natcomm2019}.
The agreement of the exponents reinforces the argument that
this nontrivial phase transition belongs to the chiral-Heisenberg universality class.

The difference between the numerical and analytical results, 
although it is still noticeable, 
is decreased as compared with the previous situation reported in Ref.~[\onlinecite{Otsuka_PRX2016}],
mainly owing to technical advances such as higher-order calculations made possible
recently~\cite{Zerf_PRD2017,Knorr_PRB2018,Janssen_PRB2014,Gracey_PRD2018}.
Among them, the four-loop renormalization-group calculation~\cite{Zerf_PRD2017} yields
results closest to the present ones with a difference of  
the order of 25\%, similar to the case of the chiral-$XY$ universality
class~\cite{Otsuka_PRB2018}.
We expect that further efforts both in the numerical and analytical approaches
will reconcile the remaining discrepancies.

Finally,
we remark on a relation between $U_{\mathrm{c}}$ and the Dirac Fermi velocity in the noninteracting limit.
In Ref.~[\onlinecite{ParisenToldin_PRB2015}],
it is suggested that
the values of $U_{\mathrm{c}}$ are well scaled 
by the geometric mean $\bar{v}$ of 
the velocities at the Dirac point. 
This is intuitively understood because, as shown in Eq.~(\ref{eq:DOS}),
$\bar{v}$ appears in the density of states $D(E)$ near the Dirac point 
in the noninteracting limit as $D(E)= |E| / (\alpha \bar{v})^2$
with $\alpha$ being a constant
$\alpha=\sqrt{\frac{V_{\mathrm{BZ}}}{2 \pi N_{\mathrm{Dirac}}}}$.
Indeed, our numerical simulations find that $U_{\mathrm{c}}$ 
scales with $\bar{v}=\sqrt{v_{\mathrm{F}}v_{\mathrm{\Delta}}}$, i.e., 
\begin{align}
 U_{\mathrm{c}} / (\alpha \bar{v}) & \simeq 2.15 & & (\Delta=1),\\
 U_{\mathrm{c}} / (\alpha \bar{v}) & \simeq 2.19 & & (\Delta=0.5).
\end{align}
In addition, we find that these scaled values are close to 
those for the honeycomb lattice model and the $\pi$-flux model.
Namely, using the values of $U_{\mathrm{c}}$ reported in Ref.~[\onlinecite{Otsuka_PRX2016}], 
we find that 
\begin{align}
 U_{\mathrm{c}} / (\alpha v_{\mathrm{F}}^{0}) &\simeq 2.33 & &(\text{honeycomb lattice model}),\\
 U_{\mathrm{c}} / (\alpha v_{\mathrm{F}}^{0}) &\simeq 2.21 & &(\text{$\pi$-flux model}),
\end{align}
where $v_{\mathrm{F}}^{0}=3t/2$ $(2t)$ is the isotropic Dirac Fermi velocity 
in the noninteracting limit,
and the value of $\alpha$ is calculated with 
$V_{\mathrm{BZ}}=\frac{8\pi^2}{3\sqrt{3}}$ $(2\pi^2)$ and 
$N_{\mathrm{Dirac}}=2$
for the honeycomb lattice model ($\pi$-flux model).
We employ $\alpha \bar{v}$, instead of $\bar{v}$ alone, 
as a measure of the energy scale, because the constant $\alpha$ 
compensates the difference among the models 
in the linear part of the density of states.
These quantitative agreements strongly support the evidence that
the effective theory for these lattice models is described by
the same model in the continuous limit, i.e., the GN model.

To conclude, 
we have revisited the quantum criticality of the phase transitions 
for the chiral-Heisenberg universality class 
in terms of the Gross-Neveu model.
The linear dispersion of the Dirac fermions is constructed by introducing 
a $d$-wave pairing field
to the Hubbard model on the square lattice.
Although the way of counting the number of fermion components is different,
the resulting effective model is the same as that of 
the widely-studied Hubbard model on the honeycomb lattice.
By exploiting large-scale quantum Monte Carlo simulations,
we have calculated the correlation ratio of the spin structure factor,
the staggered magnetization, and the quasiparticle weight.  Based 
on these quantities, 
the antiferromagnetic phase transitions have been investigated
by  several methods 
such as the crossing-point analysis and the data-collapse fit.
The conservative estimates for the critical exponents 
obtained in this work  are
$\nu=1.05(5)$, 
$\eta_{\phi}=0.75(4)$, and 
$\eta_{\psi}=0.23(4)$.
These results improve our previous estimates, 
especially for the exponent $\eta_{\phi}$, 
which is now closer to the recent independent QMC calculation~\cite{Liu_Natcomm2019}.
Indeed we have noticed that 
it is a cumbersome task to judge whether correction terms to the simple scaling ansatz 
should be included in a data-collapse fit of the Bayesian scaling analysis~\cite{Harada_PRE2011},
and our previous estimates of $\eta_{\phi}$~\cite{Otsuka_PRX2016} 
were eventually not accurate enough.
We have also shown that the anisotropy of the Dirac cones does not
affect the criticality, 
which suggests the emergent relativistic invariance at the quantum critical point.

\acknowledgements
We acknowledge
F.~F.~Assaad, T.~Sato, F.~Parisen Toldin, and Z. Wang
for valuable comments.
This work has been supported by 
Grant-in-Aid for Scientific Research from MEXT Japan 
(under Grant Nos. 18K03475, 18H01183, and 19K23433)
and by PRIN2017 MIUR prot.2017BZPKSZ. 
The numerical simulations have been performed on 
K computer provided by the RIKEN Center for Computational Science (R-CCS) 
through the HPCI System Research project 
(Project IDs: hp170162 and hp170328),
and RIKEN supercomputer system (HOKUSAI GreatWave).

\bibliography{H_BCS}

\begin{thebibliography}{83}
\bibitem [{not()}]{noteSM}%
  \BibitemOpen
  \href@noop {} {}\bibinfo {note} {In this paper, a $d$-wave BCS state with the
  Dirac quasiparticle dispersion is referred to simply as a semimetallic
  state}\BibitemShut {NoStop}%
\end{thebibliography}
\end{document}